\documentstyle[12pt]{article}
\input{epsf.tex}

\begin{document}

  \begin{flushright} \begin{small}
  DF/IST-.2001 \\gr-qc/0201162
  \end{small} \end{flushright}

\vskip 0.5cm

\begin{center}
{\bf SCALAR SYNCHROTRON RADIATION IN THE
SCHWARZSCHILD-ANTI-DE SITTER GEOMETRY} \\
\vskip 1cm
Vitor Cardoso \\
\vskip 0.3cm
{\scriptsize  CENTRA, Departamento de F\'{\i}sica,
	      Instituto Superior T\'ecnico,}\\ 
{\scriptsize  Av. Rovisco Pais 1, 1096 Lisboa, Portugal,}\\
{\scriptsize E-mail: vcardoso@fisica.ist.utl.pt}\\
\vskip 0.6cm

Jos\'e P. S. Lemos \\
\vskip 0.3cm
{\scriptsize  CENTRA, Departamento de F\'{\i}sica,
	      Instituto Superior T\'ecnico,} \\
{\scriptsize  Av. Rovisco Pais 1, 1096 Lisboa, Portugal, }\\
                      {\small\&}\\
{\scriptsize Institute of Astronomy, University of Cambridge,}\\ 
{\scriptsize Madingley Road, CB3 OHA, Cambridge, UK}\\
{\scriptsize E-mail: lemos@kelvin.ist.utl.pt}
\end{center} 
 
\bigskip

\begin{abstract}
\noindent
We present a complete relativistic analysis for the scalar radiation
emitted by a particle in circular orbit around a Schwarzschild-anti-de
Sitter black hole.  If the black hole is large, then the radiation is
concentrated in narrow angles- high multipolar distribution- i.e., the
radiation is synchrotronic.  However, small black holes exhibit a
totally different behavior: in the small black hole regime, the
radiation is concentrated in low multipoles. There is a transition
mass at $M=0.427 R$, where $R$ is the AdS radius. This behavior is 
new, it is not present in asymptotically flat spacetimes. 
\strut  
\newline 
\end{abstract}
\noindent
\section{ Introduction}
\vskip 3mm

Cyclotron and synchrotron radiations both have inherited their names
from the apparatus used to accelerate charged particles in the 30's
and 40's of last century. The radiation generated due to the charge
acceleration was a secondary effect which became a subject on itself.
Cyclotron radiation comes from electrically charged particles in a
magnetic field $\vec{B}$ with non-relativistic velocities $\vec{v}$ in
circular or spiralling (along $\vec{B}$) orbits, which spectrum
consists of a single frequency of emission, equal to the
orbital frequency of the particle in the magnetic field,
$\omega_0=2\pi\frac{q\,B}{m}$, where $q$ and $m$ are the charge and
mass of the particles, respectively. For highly relativistic particles
the spectrum is much more complex and most of the power is radiated in
a range around a frequency $\overline\omega$, 
which is many times the orbital particle
frequency $\omega_0$, $\overline{\omega}
\simeq\gamma^{3}\omega_0$, where gamma is the
Lorentz factor, into a narrow cone due to relativistic headlight
effect, and strongly linearly polarized in the plane of the circular
motion. This radiation due to relativistic circular particles 
is called synchrotron radiation.  There is of course a transition from
cyclotron to synchrotron radiation, in which the particle aquires
larger and larger velocities and higher and higher harmonics of the
fundamental mode $\omega_0$ start to be excited. Instead of receiving
a sinusoidal pulse with a sharp frequency as in the cyclotron
radiation case, the observer starts to receive a series of sharp
pulses repeating at intervals of $2\pi/\omega_0$. 
Synchrotron radiation is an important electromagnetic radiation in
astrophysical systems such as the Sun, the magnetosphere of Jupiter,
pulsars, and active galaxies (for detailed analysis of synchrotron
radiation see \cite{jackson,rybicki}).  Fields other than the
electromagnetic (vector) field, like the scalar \cite{breuerbook} and
the gravitational (tensorial) field \cite{pricesandberg}, can also radiate
synchrotonically in flat spacetime. The spectrum has for the three
fields the form $P(\omega)\propto \omega\exp(-2\omega/\omega_{\rm
crit})$, with $\omega_{\rm crit}$ being a few times $\overline\omega$.

By using the equivalence principle, one can further ask whether it is
possible for a particle in geodesic motion in a gravitational field to
emit cyclotron or synchrotron radiation.  
Indeed, cyclotron gravitational radiation has been 
studied, see, e.g., \cite{poisson}
for recent work on detailed calculations of
gravitational radiation from a particle in a geodesic circular orbit around a
Schwarzschild black hole, which is  motivated  
mainly by the possibility of detecting gravitational waves
by LIGO or VIRGO projects. 
On the other hand, Misner et al \cite{misner1,misner2}
showed that geodesic synchrotron radiation (GSR) is possible
for particles near the relativistic photonic orbits around a black
hole. This radiation was first worked out for scalar particle and
field in a Schwarzschild black hole \cite{misner2} and it showed three
main features: (i) the range of radiated frequencies $\overline
\omega$ are higher harmonics (or higher multipoles) 
of the particle's orbital frequency
$\omega_0=(M/r_0^3)^{1/3}$, namely, $\overline{\omega}\simeq \gamma^2
\omega_0$, where $\gamma$ is the Lorentz factor,
$\gamma=1/\sqrt{1-3M/r_0}$, and $M$ and $r_0$ are the mass of the
black hole and the position of the orbit, respectively, (ii) it is
beamed into narrow orbital plane angles, and (iii) it is linearly
polarized in the orbital plane \cite{breuervish}. The existence of GSR
was then studied for electromagnetic and gravitational fields
\cite{tiomno1} where it was shown that the spectrum is broader
than for the scalar field, $P(\omega)\propto
\omega^{1-s}\exp(-2\omega/\omega_{\rm crit})$, with $s$ being the spin
of the radiated field, and again $\omega_{\rm crit}$ being a few
times $\overline\omega$.  It is interesting to note, following
\cite{chitreprice}, that there are differences between ordinary (or
accelerated) synchrotron radiation (OSR) and GSR.  For OSR the
spectrum does not depend on the spin $s$ of the field, whereas for GSR
it does. This stems from the fact that geometric optics (short
wavelength approximation) is valid for a source in flat spacetime,
whereas in a strong gravitational field the effective gravitational
potential does not permit short waves within the emitting region
\cite{chitreprice}. 
GSR went into oblivion after it was shown that the GSR concept 
was not applicable astrophysically \cite{tiomno2,bardeen}, 
mainly due to the fact that it is astrophysically hard to 
put particles on photonic orbits, although some attempts 
with non-geodesic motion were made \cite{galtsov}.

However, black holes are now not only of astrophysical interest but
also are of interest to elementary particle physics. Since many
elementary particle theories predict that the vacuum is anti-de Sitter
(AdS) one should reanalyze the problem in a Schwarzschild-AdS
background and work out the similarities and the differences. This is
what we do in this paper.  In this paper we shall study a specific
important problem of radiation emission: the scalar energy emitted by
a small test particle coupled to a massless scalar field as it orbits
in circular motion a Schwarzschild-AdS black hole.  This work is then
also of interest to the AdS/CFT conjecture \cite{maldacena1}, which in
turn has attracted much attention to the investigation of
asymptotically AdS spacetimes.  According to it phenomena in the bulk
can be reinterpreted as phenomena in the Conformal Field Theory (CFT)
boundary. In addition, the black hole corresponds to a thermal state
in the conformal field theory, and the decay of a test field in the
black hole spacetime, correlates to the decay of the perturbed state
in the CFT.  One knows for example that the decay of this test field
is correctly described by the so called quasinormal modes, which have
been recently computed for AdS spacetimes (see,
e.g. \cite{hubenycardoso}, and references therein).  In order to gain
deeper insight into the conjecture, one needs to understand how the
information about the bulk is encoded in the boundary, namely by
probing the bulk with test fields, or pointlike particles
\cite{holography}, and to understand how the spacetime answers to
specific perturbations.  A specific problem, the radial infall of a
small test particle into a Schwarzschild-AdS black hole and consequent
emission of radiation is being addressed  \cite{cardosolemos1}, where
it is found that the signal is dominated by
quasinormal ringing (for the analogous problem in the BTZ 3-dimensional 
black hole see \cite{cardosolemos2}). 

The paper is organized as follows. In section 2 , we introduce 
the problem, and the basic mathematical apparatus needed to solve it.
In section 3 , we present the numerical results obtained, and the
most important features of these numerical results.
In section 4 , we present some concluding remarks.

\noindent
\section{Equations and Formalism}
\vskip 3mm
\noindent
\subsection{The problem}
\vskip 3mm
Black holes in AdS spacetimes in several dimensions have been recently
study.  All dimensions up to eleven are of interest in superstring
theory, but experiment singles out four dimensions (4D) as the most
important. In 4D general relativity, an effective
gravity theory in an appropriate string theory limit, the Kerr-Newman
family of four-dimensional black holes can be extended to include a
negative cosmological constant \cite{bcarter}. Our 
background is the 4D Schwarzschild-AdS black hole metric,
\begin{equation}
ds^{2}= f(r) dt^{2}- \frac{dr^{2}}{f(r)}-
r^{2}(d\theta^{2}+\sin^2\theta d\phi^{2})\,,
\label{lineelement}
\end{equation}
where, $f(r)=(\frac{ r^{2}}{R^2}+1-\frac{2M}{r})$,
$R$ is the AdS radius and $M$ the black hole mass.  
We now
consider a small particle coupled to a massless scalar field,
described by the interaction action
\cite{misner2} 
\begin{equation}
S=-\frac{1}{8 \pi} \int g^{1/2} \varphi _{;a} \varphi ^{;a} d^4x-
 m_0 \int (1+q_s \varphi)(-g_{ab}\dot{z}^a \dot{z}^b)^{\frac{1}{2}} 
d\lambda \,,
\label{action}
\end{equation}
where $q_s$ is the scalar charge carried by the test particle, and
$m_0$ its mass.  The mass $m_0$ is supposed to be small and the scalar
field treated as a perturbation, in the sense that the background
metric is still given by (\ref{lineelement}). This means that the
particle travels on a geodesic of the spacetime.

After the usual decomposition in spherical harmonics,
$\varphi(r,\omega,\theta,\phi)=\frac{1}{r}\sum_{lm}
\psi(r,\omega) Y_{lm}$,
where $Y_{lm}$ are the usual
spherical harmonics,
and a Fourier transform
$\Psi(r,\omega)=\frac{1}{(2\pi)^{1/2}} \int_{-\infty}^{\infty} 
e^{i \omega t} \psi(r,t)$,
the evolution of the scalar field is given by the wave equation with 
a generic source term $S$:
\begin{equation}
\frac{\partial^{2} \Psi(r)}{\partial r_*^{2}} +
\left\lbrack\omega^2-V(r)\right\rbrack\Psi(r)=S \,.
\label{waveequation1}
\end{equation}
Here,
\begin{equation}
S=\int 
\frac{1}{(2 \pi)^{1/2}}\frac{f(r)}{\gamma r}Y_{lm}(\theta,\phi)\,e^{i\omega t}
\,\delta(r-r(t))dt\,, \label{source1}
\end{equation}
where $r(t)$, $\theta(t)$ and $\phi(t)$ are the spatial Schwarzschild 
coordinates of the test particle, and $\gamma=1/\sqrt{1-3M/r}$.
The potential $V$ appearing in equation (\ref{waveequation1}) is given by
\begin{equation}
V(r)=f(r)\left\lbrack2+\frac{2M}{r^3}
+
\frac{l(l+1)}{r^2}\right\rbrack \,.
\label{potential}
\end{equation}
The tortoise coordinate $r_*$ is defined as
$\frac{\partial r}{\partial r_*}= f(r)$.
\noindent
\subsection{Circular geodesics in the Schwarzschil-AdS geometry}
\vskip 3mm

Since to the best of our knowledge, no full investigation has been made
on the geodesics in this spacetime, we shall study the circular null
and timelike geodesics in the Schwarzschild-AdS geometry,
which will be useful in what follows.
The integrals of motion
$f dt/d\tau=E$, and $ r^2d\phi/d\tau=L$, where $E$ is an energy
parameter and $L$ an angular momentum parameter,
plus the constancy of the Lagrangian yield

\begin{equation}
\dot{r}^2+V(r)^2=E\,;\,\,V(r)^2=f(\epsilon+L^2/r^2)\,,
\label{principalequationgeodesic}
\end{equation}
where $\epsilon=0$ for null geodesics and $\epsilon=1$ for
timelike geodesics.

(i) Timelike circular geodesics:
in this case $\epsilon=1$. Demanding $dV^2/dr=0$ we get
\begin{equation} 
L^2=\frac{r^5/R^2+Mr^2}{r-3M}\,,
\label{timelike1}
\end{equation}
and
\begin{equation}
\omega_0 \equiv d\phi/dt = (\frac{1}{R^2}+\frac{M}{r^3})^{1/2}.
\label{timelike2}
\end{equation}
This means that circular timelike geodesics may exist for any $3M<r<\infty$.
Let us now see which of them are stable and which are
unstable.  If we differentiate twice the potential and then substitute
$L^2$ given by (\ref{timelike1}), we obtain
\begin{equation}
d^2V^2/dr^2=-2\frac{15Mr^3-4r^4+6M^2R^2-MR^2r}{r^3R^2(r-3M)}.
\label{timelike3}
\end{equation}
Now, (\ref{timelike3}) has one and only one real root in $3M<r<\infty$.
This root will give us the range of allowed
stable and unstable circular orbits. The roots can be done numerically,
but two important limiting cases can be studied analytically, and these are 
the very large and very small black hole limit.
For very large black holes, we can immediatly see that the root is at
$r=15/4M=3.75M$.
For very small black holes we can see that the root is at $r=6M$, which 
is what we expect: for small black holes the results of the Schwarzschild
geometry should carry over.
For any intermediate mass, the last stable circular orbit
lies in between $r=3.75M$ and $r=6M$.

(ii) Null circular geodesics: in this case we require $\epsilon=0$, $\dot{r}=0$ and
$dV^2/dr=0$, which gives us
\begin{equation}
r=3M\,;\,\, \frac{E^2}{L^2}=\frac{1}{R^2}+\frac{1}{27M^2}\,,
\label{nullgeodesics}
\end{equation}
and $d^2V^2/dr<0$. Thus, just as in Schwarzschild spacetime, a circular
orbit of radius $3M$ is the only allowed null geodesic, which is furthermore
an unstable one.

\noindent
\subsection{ The Green's function solution}
\vskip 3mm
Under these conditions, (\ref{waveequation1}) becomes
\begin{equation}
\frac{\partial^{2} \Psi(r)}{\partial r_*^{2}} +
\left\lbrack m^2\omega_0^2-V(r)\right\rbrack\Psi(r)=S_l \delta(r-r_0) \,,
\label{waveequation2}
\end{equation}
with $S_l=4f(r)\pi Y_{lm}(\pi/2,0)/(\gamma r)$, and again 
$\gamma={1}/\sqrt{1-3M/r_0}$.
We would like to draw attention to the fact that for
circular geodesic motion the frequency $\omega$ must
be a multiple of the frequency $\omega_0$ of revolution
around the black hole, i.e., $\omega=m\omega_0$.
In (\ref{waveequation2}) we have rescaled $r$, 
$r\rightarrow\frac{r}{R}$, and measure 
everything in terms
of $R$, i.e., $\omega$ is to be read $\omega R$, $\Psi$ is to be 
read $\frac{R}{q_s m_0}\Psi$ and
$r_+$, the horizon radius is to be read $\frac{r_+}{R}$. 
Equation (\ref{waveequation2}) is to be solved under 
the boundary conditions appropriate for
Schwarzschild-AdS black holes:  ingoing waves at the
horizon ($\Psi \sim A e^{-iwr_*}$), and reflective boundary conditions
($\Psi=0$) at infinity \cite{avis}. Of course, under these conditions, all
the energy eventually goes down the black hole, and this is the energy
we are interested in compute.
To implement a numerical solution, we note that two independent solutions 
$\Psi_1$ and $\Psi_2$
of (\ref{waveequation2}),
with the source term set to zero, have the behavior:
\begin{eqnarray}
\Psi_1 \sim e^{-i\omega r_*}\,,r \rightarrow r_+ \\
\Psi_1 \sim Ar +B/r^2\,,r \rightarrow \infty \\
\Psi_2 \sim 1/r^2\,,r \rightarrow \infty \\
\Psi_2 \sim Ce^{i\omega r_*}+De^{-i\omega r_*}\,,r \rightarrow r_+ 
\label{behavior}
\end{eqnarray}
The Wronskian of these two solutions is $W=2Ci\omega$.
By a standard Green's function analysis, we get that the solution to the 
inhomogeneous
equation (\ref{waveequation2}) behaves, near the horizon, as
\begin{eqnarray}
\Psi= \frac{e^{-i\omega r_*}}{(2i\omega C)} 
\int_{r_+}^{\infty} \frac{S\Psi_2}{f}\, \delta(r-r_0)dr\\
    =  \frac{e^{-i\omega r_*}}{(2i\omega C)} \frac{S}{f}(r_0)\Psi_2(r_0).
\label{solution}
\end{eqnarray}
All we need to do is find a solution $\Psi_2$ of the 
corresponding homogeneous equation satisfying
the above mentioned boundary conditions (\ref{behavior}).
In the numerical work, we chose to adopt $r$ as the independent variable,
therefore avoiding the numerical inversion of $r_*(r)$.
The integration was started at a large value of $r=r_i$, 
which was $r_i=10^5 r_+$ typically.
Equation (\ref{behavior}) was used to infer the boundary 
conditions $\Psi_2(r_i)$ and $\Psi_2'(r_i)$.
We then integrated inward from $r=r_i$ into typically $r=r_++10^{-6}r_+$.
Equation (\ref{behavior}) was then used to get $C$.

The total power $P_{\rm tot}$ radiated into the black hole is
$
P_{\rm tot}=\sum_{l,m >0} \frac{\omega^2}{2\pi}|\Psi|^2\,, 
$
and the power spectra $P$ 
(power radiated at a given frequency $\omega=m\,\omega_0$) is
\begin{equation}
P(m\omega_0)=\sum_{l \geq |m|} \frac{(m\omega_0)^2}{2\pi}|\Psi|^2\,
\label{powerspectra}
\end{equation}
where $m$ is fixed in the summation.

\noindent
\section{Numerical results}
\noindent
\subsection{ Large and intermediate black holes}
\vskip 3mm
We define large and intermediate black holes as 
black holes with $r_+ \geq1$. 
The results of the numerical integration are shown in Figures 1a and 1b
for orbits at $r_0=5$ and $r_0=20$, respectively, 
and a horizon radius $r_+=1$.
In Figure 2 it is shown the power spectra as a 
function of the azimuthal quantum
number $m$ for large $m$ in a semi-log plot.

\centerline{\epsffile{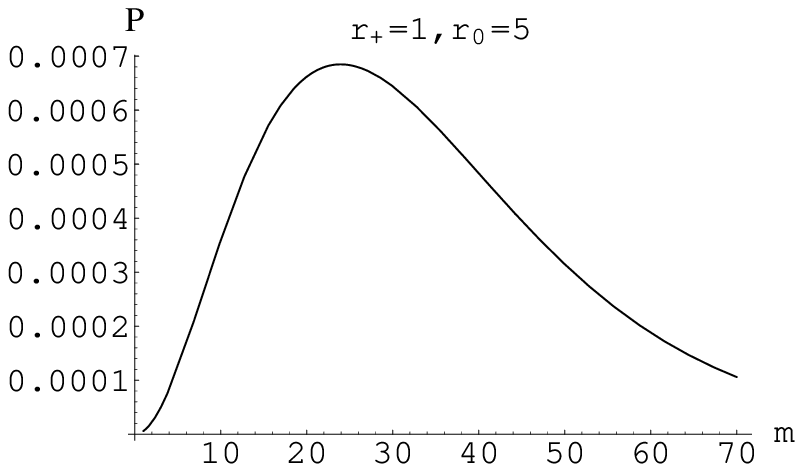}}
{\noindent {\small Figure 1a. Scalar radiation power spectra as a function of
the azimuthal quantum number $m$, for an orbit with $r_0=5$,
around a Schwarzschild-AdS black hole with $r_+=1$.}
\vskip1mm
\centerline{\epsffile{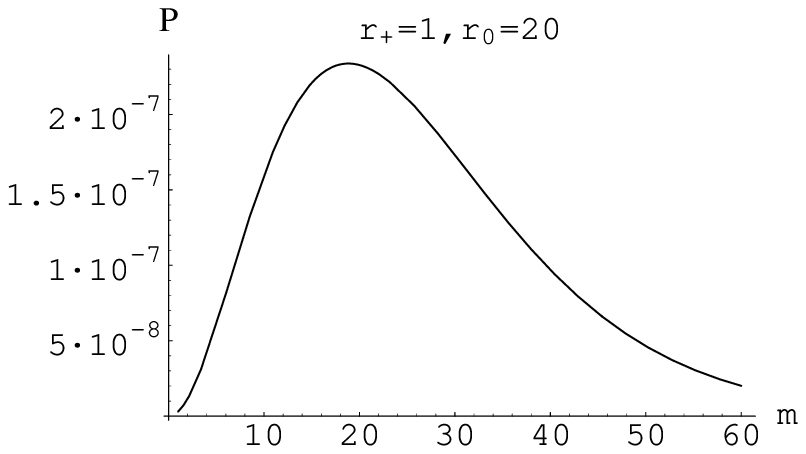}}
{\noindent {\small Figure 1b. Scalar radiation power spectra 
as a function of the angular quantum
number $m$, for an orbit with  $r_0=20$, around a 
Schwarzschild-AdS black hole with $r_+=1$ .

\vskip3mm
\centerline{\epsffile{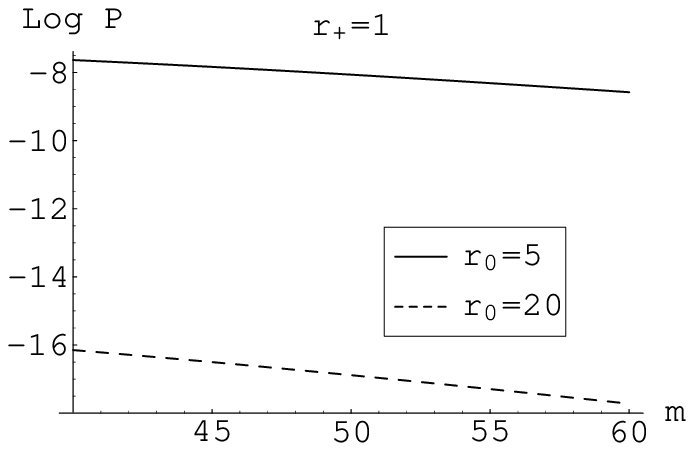}}
{\noindent {\small Figure 2. Plot of $\log P\times m$ for large $m$. 
One can see that $\log P$ is a linear
function of $m$, so that for large $m$ the power 
output decreases exponentially. 
\bigskip

The results for the large black hole regime are the following:

\noindent (1) The power radiated in the $l=m$ modes is more than 95\%
of the total power.

\noindent (2) High multipoles, as one can see from Figs. 1a and 1b,
are clearly enhanced, and the radiation is synchrotronic. To
see that most of the radiation is in fact confined to small angles, we
have ploted the angular distribution of the power in Fig. 3.

\noindent (3) The location of the peak of the spectrum increases with
increasing radius of the circular orbit, so that truly synchrotronic
radiation occurs only for highly relativistic (unstable) orbits. 

\noindent (4) From Figs. 1a and 1b, and as expected, we see that the
power output decreases with increasing $r_0$, the orbit radius. In
fact one can easily prove that for high $r_0$ the power goes as
$1/r_0^6$.

\noindent (5) The location of the peak increases dramatically with the mass of the
hole. So for large black holes the emission is dominated by very large
$m$. This strong dependence of the location of the peak on the black
hole mass, makes us believe that small black holes do not emit
synchrotronic radiation. This will be seen to be true in the next
subsection.

\noindent (6) Furthermore, the power decays as an exponential power of
the frequency, for high frequencies, which is evident from Fig. 2,
where we show a plot of $\log P\times m$; An analytical approximation
for large $m$ seems very difficult to achieve, due to the behavior of
the potential. We find numerically that for large $m$, the exponential 
dependence of the power spectra is 
\begin{equation}
P \sim e^{-\frac{2\omega}{\omega_{\rm crit}}}
\label{behaviorlargellargeM}
\end{equation}
where $\omega_{\rm crit}=\frac{2\,\gamma\,M^2\,\omega_0}{0.09}$. 
We conclude from this that most of the radiation is emitted
at a given $\overline\omega$ and that $\overline\omega$ increases
with the mass of the black hole 
(numerically we have found $\overline\omega \sim M^2$).

\vskip3mm
\centerline{\epsffile{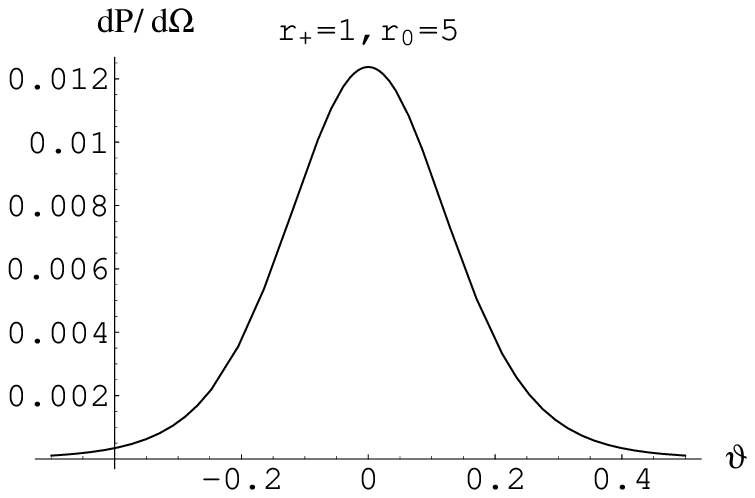}}
{\noindent {\small Figure 3. Scalar power per unit solid angle 
$\vartheta \equiv \pi/2-\theta$, for the case of a scalar particle 
orbiting around a $r_+=1$ black hole, with an orbital radius $r_0=5$.
\bigskip

\noindent
\subsection{ Small black holes}
\vskip 3mm

In Figs. 4a and 4b we show the numerical results for small black holes, 
$r_+<1$. 


\centerline{\epsffile{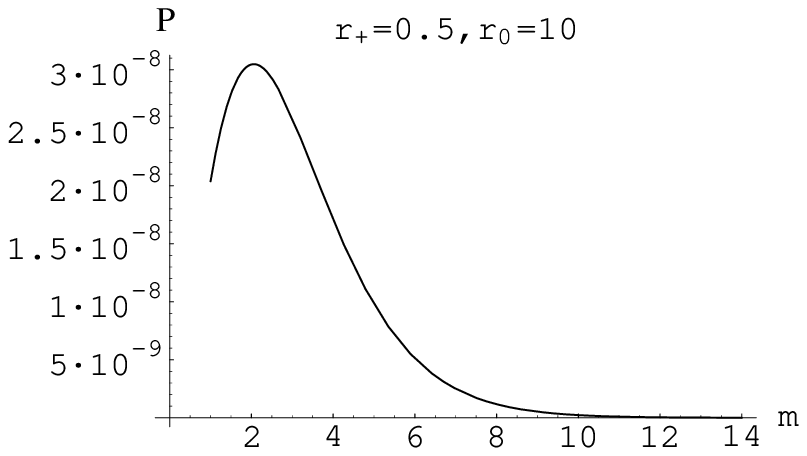}}
{\noindent {\small Figure 4a. Scalar radiation power as a function of
the angular quantum number $m$, for an orbit with $r_0=10$,
around a Schwarzschild-AdS black hole with $r_+=0.5$.}
\vskip1mm

\centerline{\epsffile{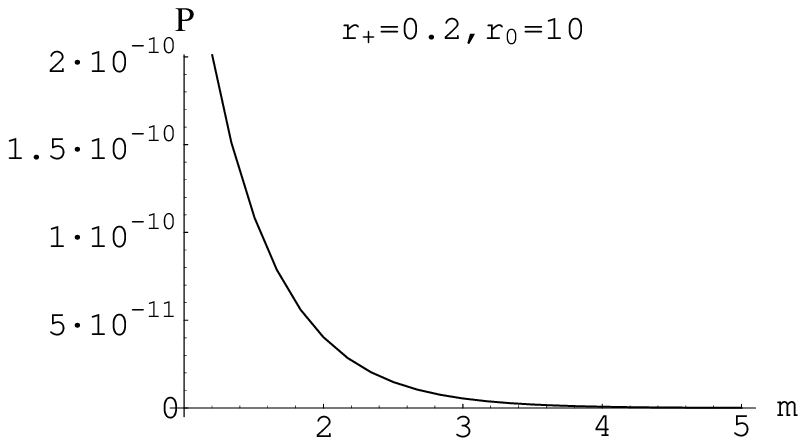}}
{\noindent {\small Figure 4b. Scalar radiation power 
as a function of the angular quantum
number $m$, for an orbit with  $r_0=10$, around a 
Schwarzschild-AdS black hole with $r_+=0.2$.}

\medskip
In the small black hole regime, we can see a completely different behavior, 
in that now the power decreases
monotonically with $m$, the angular quantum number.
Numerically, we find that, independently of $r_0$, the location (in $l$)
of the peak of the spectrum progressively approaches $m=0$ as one lowers 
the mass of the black hole. Finally, for $M<0.427 R$ ($r_+ <0.618 R$) 
the spectrum is monotonically decreasing in $m$, for all $m$.
Still, for large $m$ the spectrum continues to decay exponentially.

\noindent
\subsection{Conclusions}
\vskip 3mm
We have computed the scalar radiation emitted by a scalar test particle 
moving in a geodesic circular orbit around a Schwarzschild-AdS 
black hole. For large black holes, the radiation is confined to small
angles, and we therefore have what can be called scalar synchrotron radiation.
However, the spectrum depends drastically on the size of the black hole.
For black holes with masses $M<0.427 R$ the spectrum does not have a peak in $m$,
and so in this regime there is no synchrotron radiation.
One might be tempted at first sight to say that small black holes in AdS 
spacetime should behave like Schwarzschild black holes. Our numerical results
show that this is not true: the boundary
conditions at infinity have changed.

These results, plus results on all previously mentioned works on AdS 
spacetime, allows us to slowly start building a small catalog of 
dynamical processes in AdS spacetimes.
It's worth emphasizing that this is extremely important if one wants
to fully understand the AdS/CFT duality and whatever surprises
there may be to unfold in AdS spaces.
Indeed here, in relation to the AdS/CFT correspondence, one can 
say that to the black hole corresponds a thermal
bath, to the orbiting particle  corresponds a travelling soliton 
(lump of energy), and to
the scalar field waves correspond particles decaying into bosons of
the associate operator of the gauge theory. Thus in the CFT boundary
one has a travelling soliton perturbing the thermal sate and irradiating 
particle pairs \cite{maldacena2}.

\bigskip
\section*{Acknowledgments} This work was partially funded
by Funda\c c\~ao para a  Ci\^encia e Tecnologia (FCT) 
through project SAPIENS 36280. VC  also 
acknowledges finantial support from FCT 
through PRAXIS XXI programme.
JPSL thanks Observat\'orio Nacional do Rio de Janeiro for
hospitality. We thank conversations with Donald Lynden-Bell. 


\end{document}